\documentclass[twocolumn,a4paper,showpacs,preprintnumbers,amsmath,amssymb,nofootinbib,floatfix]{revtex4}
\input psfig.sty  
\def \be {\begin{equation}} 

\def \ee {\end{equation}} 
\def \bea {\begin{eqnarray}} 
\def \eea {\end{eqnarray}} 

\usepackage{graphicx}
\usepackage{dcolumn}
\usepackage{bm}
\usepackage{epsfig}

\begin{document}

\title{An estimate of the dark matter density from galaxy clusters and supernovae data}
\author{R. F. L. Holanda$^{1,2}$} \email{holandarfl@fisica.ufrn.br}
\author{R. S. Gon\c{c}alves$^{3,4}$} \email{rsousa@on.br}
\author{J. E. Gonzalez$^{1}$}\email{jegonzalez@fisica.ufrn.br}
\author{J. S. Alcaniz$^{4,1}$ } \email{alcaniz@on.br}

\affiliation{\\$^1$Departamento de F\'{\i}sica, Universidade Federal do Rio Grande do Norte, 59072-970, Natal - RN, Brasil,\\
$^2$Departamento de F\'{\i}sica, Universidade Federal de Sergipe, 58429-900, S\~ao Crist\'ov\~ao - SE, Brasil,\\
$^3$Departamento de F\'{\i}sica, Universidade Federal do Maranh\~ao, 59300-000, S\~ao Luiz - MA, Brasil,\\
$^4$Observat\'orio Nacional, 20921-400, Rio de Janeiro - RJ, Brasil}

\date{\today}

\begin{abstract}
In this paper, we discuss a model-independent way to obtain the present dark matter density parameter ($\Omega_{\rm{c,0}}$) by combining gas mass fraction measurements in galaxy clusters ($f_{gas}$), type Ia supernovae  (SNe Ia) observations and measurements of the cosmic baryon abundance from observations of absorption systems at high redshifts. Our estimate is $\Omega_{\rm{c,0}} = 0.244 \pm 0.013$ ($1\sigma$). By considering the latest local measurement of the Hubble constant, we obtain  $\Omega_{\rm{M,0}} = 0.285 \pm 0.013$ ($1\sigma$) for the total matter density parameter. 
We also investigate departures of the evolution of the dark matter density with respect to the usual $a^{-3}$ scaling, as usual in interacting models of dark matter and dark energy. As the current data cannot confirm or rule out such an interaction, we perform a forecast analysis to estimate the necessary improvements in number and accuracy of upcoming $f_{gas}$ and SNe Ia observations to detect a possible non-minimal coupling in the cosmological dark sector. 
  
\end{abstract}
\maketitle

\section{Introduction}

Even after two decades of the discovery of the cosmic acceleration, the physical mechanism behind this phenomenon remains unknown~\cite{Sahni:1999gb,Peebles:2002gy,Padmanabhan:2002ji,Weinberg:2012es}. As is well known, within the framework of the general theory of relativity, this result implies the existence of a dark energy component, an exotic field with fine-tuned properties that drives the late-time cosmic expansion\footnote{Another possibility is that the matter content of the universe is subject to dissipative processes (see e.g.~\cite{Lima:1999rt})}.  In this context, the standard $\Lambda$ - Cold Dark Matter ($\Lambda$CDM) model is a great triumph of observational cosmology, as it provides a good fit to a wide range of data with just six free parameters\footnote{Currently, the $\Lambda$CDM model faces the so-called $H_0$ tension problem, a $4\sigma$-discrepancy between local and model-independent measurements of the Hubble constant~\cite{Riess:2019cxk}, $H_0$, and estimates of this quantify obtained in the context of the model using CMB data~\cite{Aghanim:2018eyx}.}. However, at the same time,  the model presents some puzzles for theorists, such as the very nature of the dark matter and dark energy as well as fine-tuning and coincidence problems~\cite{Weinberg:1988cp,Weinberg:2000yb}. 

In order to alleviate these problems,  a number of alternative cosmologies have also been proposed (see e.g. \cite{Dvali:2000hr,Capozziello:2002rd,Alcaniz:2002qh,Sahni:2002dx,Dev:2003cx,Carvalho:2008am}). This scenario makes clear that model-independent measurements of cosmological quantities, which is a challenge for observational cosmology, is of fundamental importance for a proper evaluation of the possibilities (see \cite{Heavens:2014rja,Carvalho:2015ica} for a discussion). Among these possibilities, an additional, non-gravitational  interaction between the dark matter and dark energy fields has been largely explored in the literature~\cite{Ozer:1985ws,Amendola:1999er,Barrow:2006hia,CalderaCabral:2008bx,Costa:2009wv,Alcaniz:2012mh}. This kind of approach may in principle provide a mechanism to alleviate the coincidence problem, being also consistent with current observational data measuring the expansion rate and clustering of matter in the Universe~\cite{Alcaniz:2012mh}.  It is well true that no observational evidence of such interaction has so far been unambiguously presented, although a weak coupling in the dark sector cannot yet be  excluded. For instance, in a recent analysis~\cite{alcaniz},  the cosmic expansion history was reconstructed using model-independent Machine Learning techniques, showing  that while the standard evolution is consistent with the currently available background data at $3\sigma$ confidence level, some deviations from the standard cosmology are present at low redshifts, which may be associated with the $H_0$-tension problem in the context of interacting models (see also \cite{DiValentino:2017iww,Carneiro:2018xwq}). From these and other recent results, it is clear that further investigation is needed to better understand features of the interaction in the dark sector.

\begin{figure*} \label{fig1}
\centering
\includegraphics[width=0.47\textwidth]{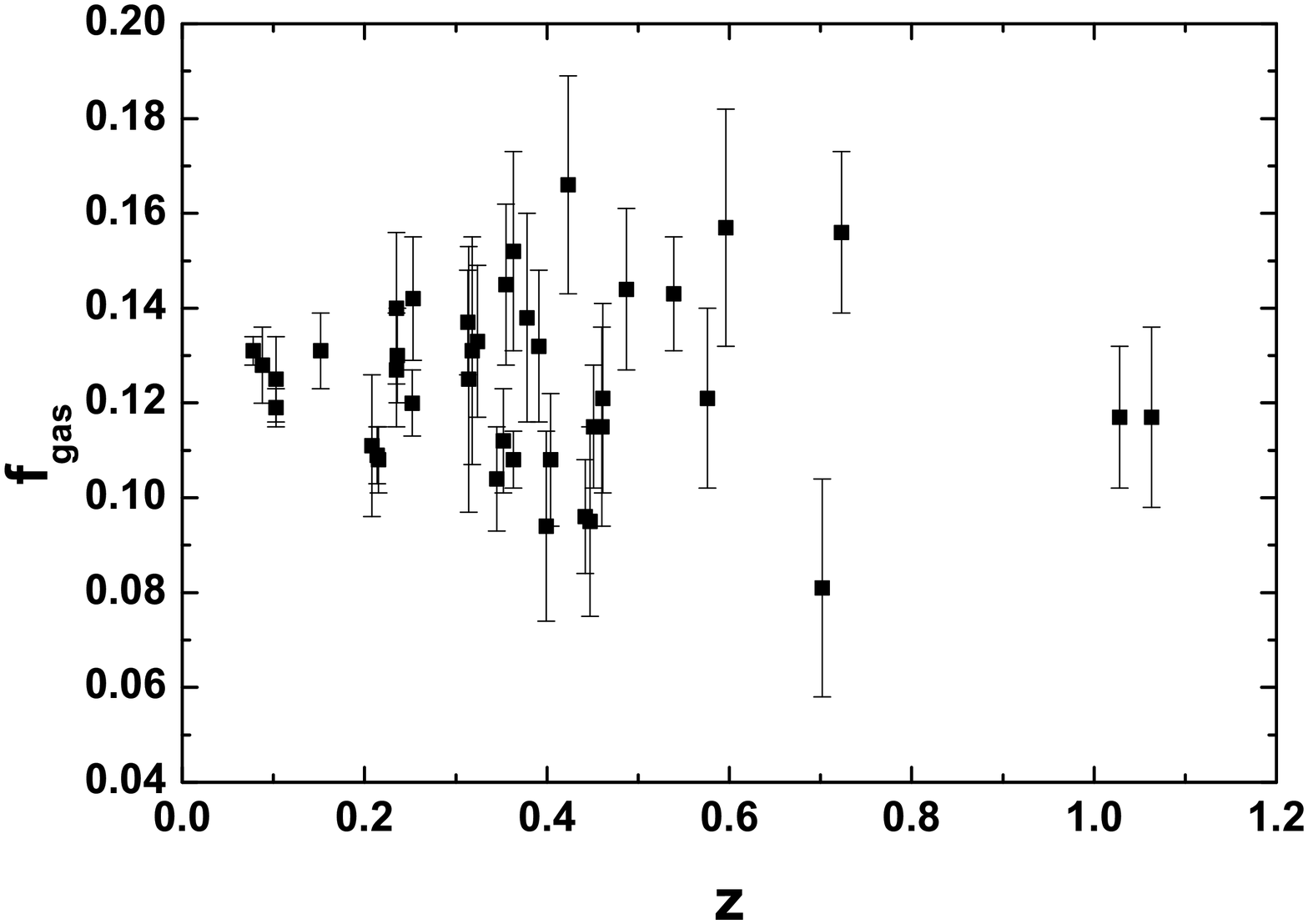}
\includegraphics[width=0.47\textwidth]{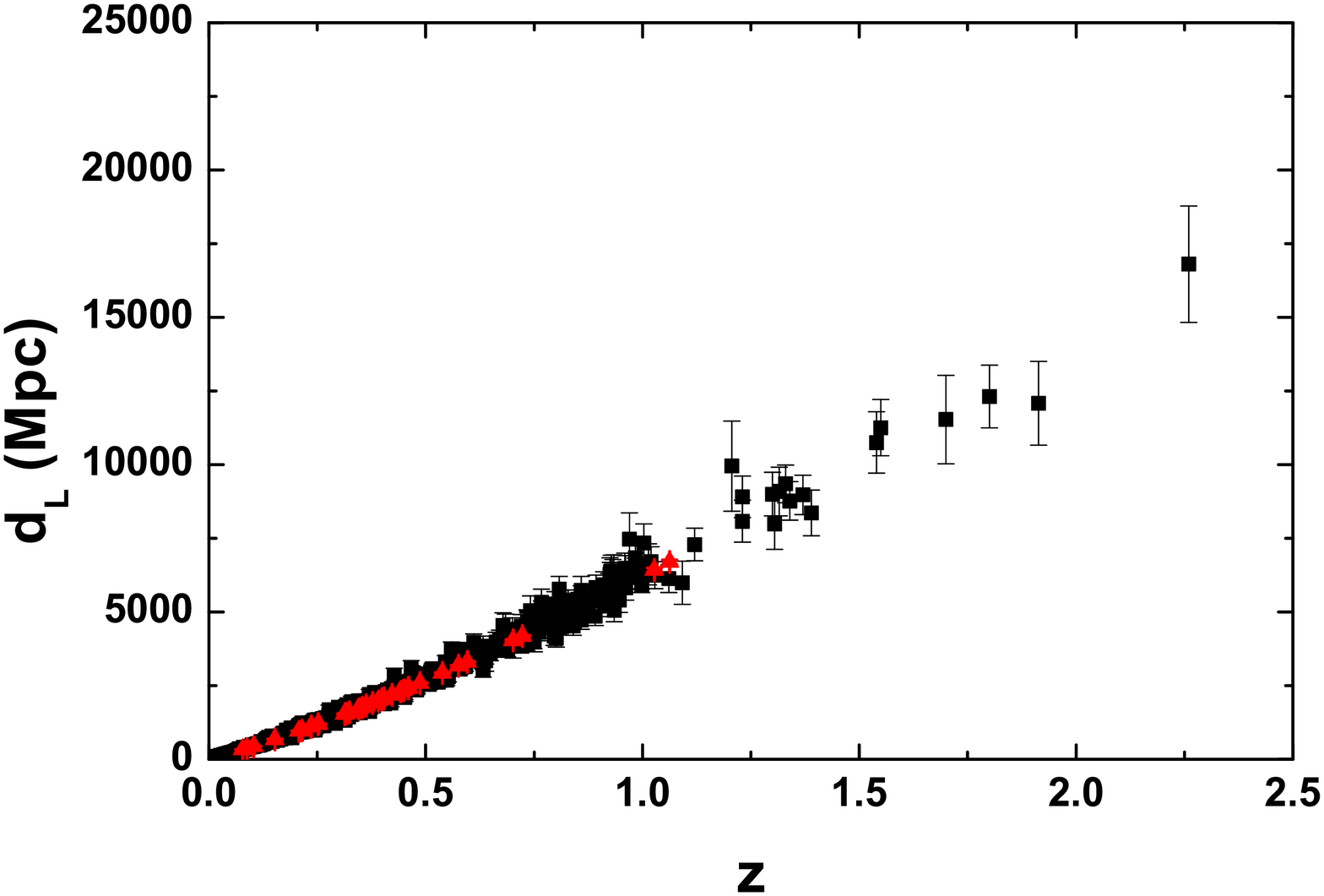}
\caption{{\it{Left)}} Measurements of $f_{gas}$~\cite{Mantz2014} used in our analysis. {\it{Right)}} The black squares show the luminosity distance as a function of the redshift obtained from the Pantheon SNe Ia compilation~\cite{pan}. On the other hand, by taking into account the full covariance matrix of the SNe (statistics and systematic), the luminosity distance for each galaxy cluster is obtained by  applying  the GP reconstruction method (red triangles). } 
\end{figure*}

{  The goal of this paper is twofold: first, we propose a model-independent way to estimate the present dark matter density parameter, $\Omega_{\rm{c,0}}$. We use 40 gas mass fraction measurements ($f_{gas}$) in galaxy clusters lying in the redshift range $0.078 \leq z \leq 1.063$~\cite{Mantz2014}, the most recent  type Ia supernovae (SNe Ia) compilation~\cite{pan}, and measurements of the cosmic baryon abundance from observations of absorption systems at high redshifts~\cite{Cooke2017}. We also discuss constraints on a possible departure from the standard dark matter evolution law, $\rho_{\rm{c}}(a) \propto a^{-3}$, by inserting the standard evolution in a more general framework, 
$\rho_{\rm{c}}(a) \propto a^{-3+\epsilon(a)}$ \cite{wang2,alcaniz2}, which arises from a non-gravitational  interaction between the dark fluids. Second, we  perform a forecast analysis using Monte Carlo simulations to discuss the necessary improvements in number and accuracy of upcoming $f_{gas}$ and SNe Ia observations to detect possible departures from the standard case $\epsilon = 0$. }

This paper is organized as follows. In Section 2, we describe the method here proposed  whereas in Section 3 we briefly describe the observational data used in our analysis. The statistical analysis and the estimates of the dark matter density are presented and discussed in Section 4. In Section 5 we present the results of our simulations. Section 6 summarizes our main conclusions.

\section{Methodology}

The main observational quantity used in our analysis is the gas mass fraction ($f_{gas}^{X-ray}(z)$) obtained from X-ray measurements in galaxy clusters. Mathematically, it can be described as~\cite{Mantz2014,Ettori2004,Ettori2006,Ettori2009,Allen2008,Goncalves:2011ha}
\begin{eqnarray}
\label{EqFgas}
  f_{gas}^{X-ray}(z) = A(z) K(z) \,  \gamma(z) \left( \frac{\Omega_{\rm{b}}(z)}{\Omega_{\rm{M}}(z)} \right) \left[ \frac{d_A^\mathrm{fid}(z)}{d_A(z)} \right]^{3/2}, \;
\end{eqnarray}
where $\Omega_{\rm{M}}(z)$ is the total mass density parameter, which corresponds to the sum of the baryonic mass density parameter, $\Omega_{\rm{b}}(z)$, and the dark matter density parameter, $\Omega_{\rm{c}}(z)$. The $A(z)$ factor stands for the angular correction factor, which is close to unity for all cosmologies and redshifts of interest and can be neglected without significant loss of accuracy \cite{Mantz2014,Allen2008}. The term in brackets corrects the angular diameter distance $d_A(z)$ from the fiducial model used in the observations, $d_A^\mathrm{fid}(z)$, which makes these measurements model-independent. 

{  The parameters $\gamma(z)$ and $K(z)$  correspond, respectively, to the depletion factor, i.e., the rate by which the hot gas fraction measured in a galaxy cluster is depleted with respect to the baryon fraction universal mean and to the bias of X-ray hydrostatic masses due to both astrophysical and instrumental sources. We adopt the value of $\gamma=0.848 \pm 0.085$ in our analysis, which was obtained from hydrodynamical simulations \cite{Planelles2013} (see also a detailed discussion in section 4.2 of the Ref.~\cite{Mantz2014}). The $\gamma$ parameter has also been estimated via observational data (SNe Ia, gas mass fraction, Hubble parameter) with the values found in full agreement with those from hydrodynamical simulations (see \cite{Holandajcap2017,zang2019}). Finally, for the parameter $K(z)$,  we use the value reported in~\cite{apple} in which {\it Chandra} hydrostatic masses to relaxed clusters were calibrated with accurate weak lensing measurements from the Weighing the Giants project. The $K(z)$ parameter was estimated to be $K=0.96 \pm 0.09 \pm 0.09$ (1$\sigma$ statistical plus systematic errors) and no significant trends with mass, redshift or the morphological indicators were verified. It is worth mentioning that these two parameters have a mild dependence on the cosmological background
 \cite{Planelles2013}.}
\begin{figure}
\centering
\includegraphics[width=0.5\textwidth]{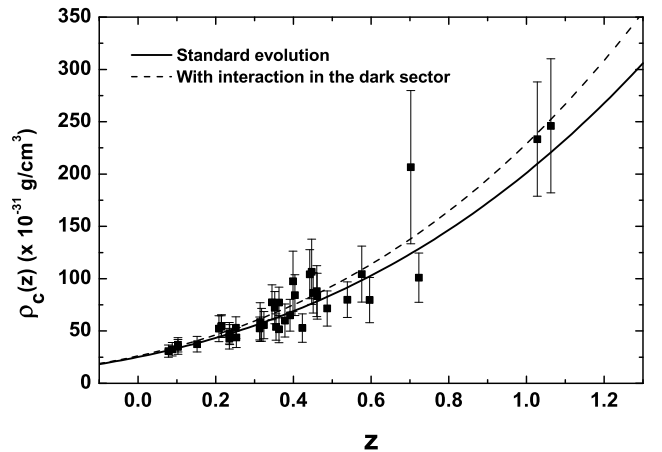}
\caption{ The evolution of the dark matter density estimated from the observational data of Fig. 1 using Eq. (\ref{RhoObs}).}  
\end{figure}

\begin{figure*}
\centering
\includegraphics[width=0.47\textwidth]{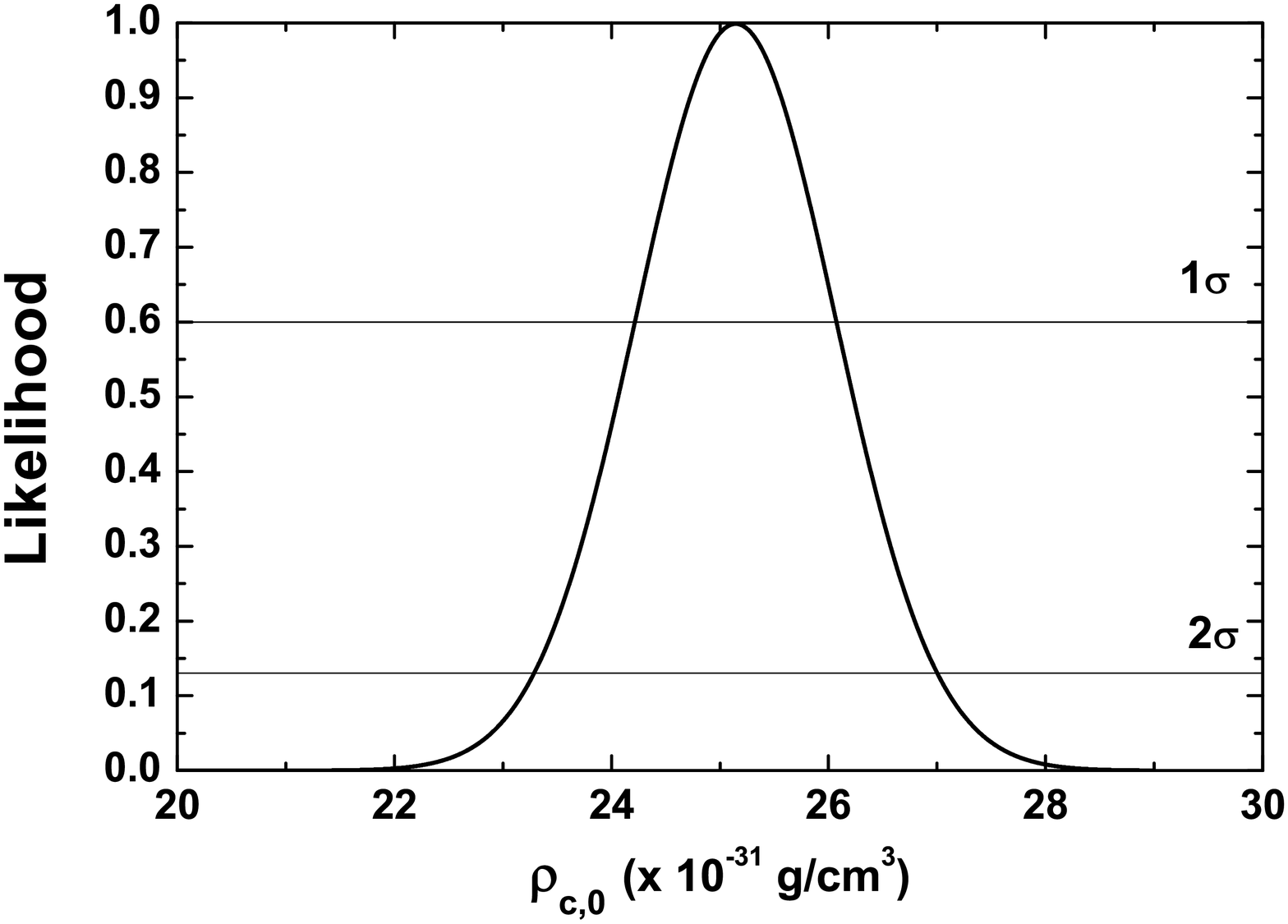}
\includegraphics[width=0.47\textwidth]{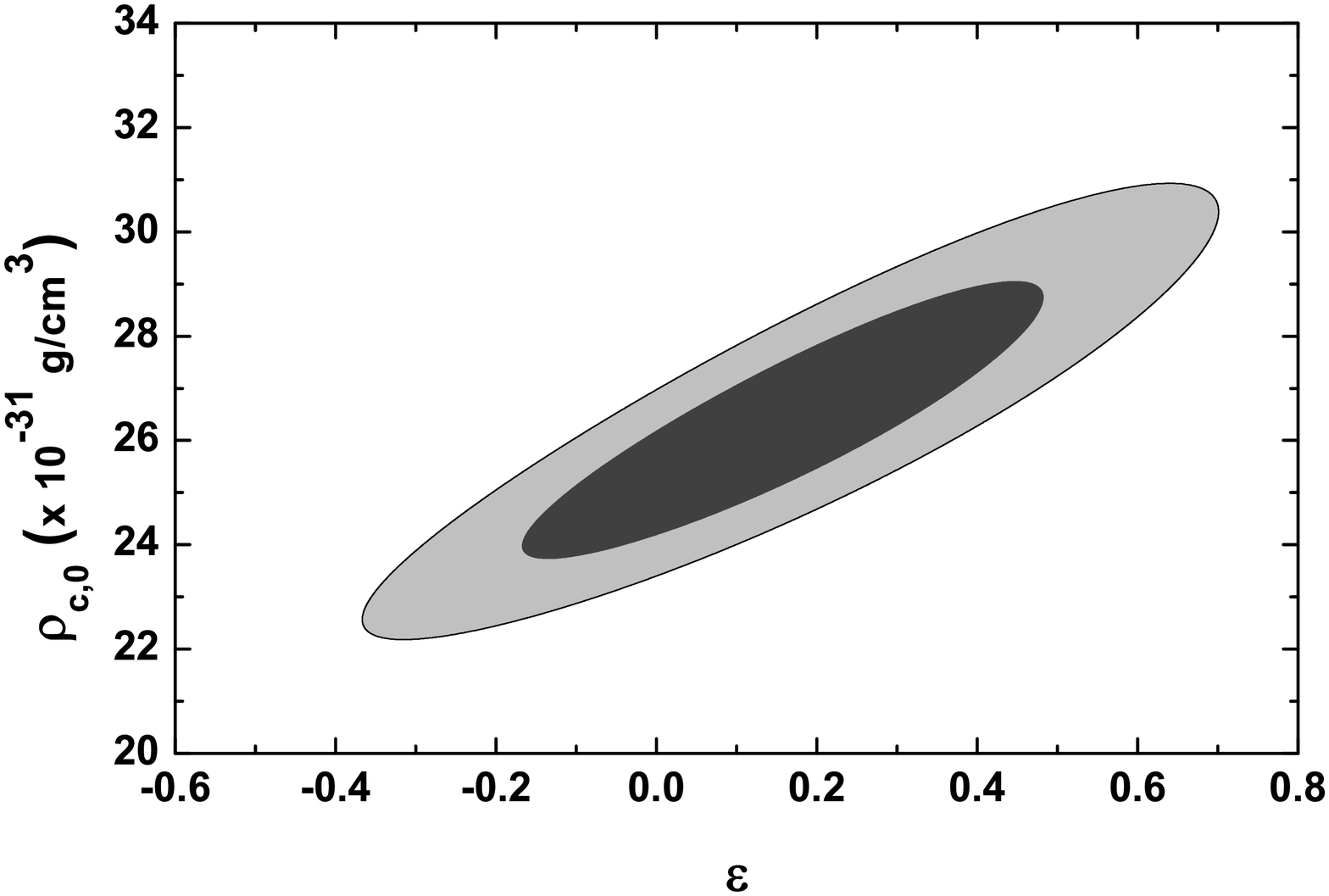}
\caption{The results of our statistical analysis. {\it{Left)}}  Likelihood of $\rho_{\rm{c,0}}$. The horizontal lines correspond to the observational best fit, $1\sigma$ and $2\sigma$ respectively. {\it{Right)}} Contour region for the plane $\rho_{DM,0}$ -- $\epsilon$. The analysis shows a good agreement with the standard evolution of the dark matter density ($\epsilon = 0$).}
\end{figure*}

The density parameter of a given component $i$ is defined as $\Omega_i=\rho_i/\rho_{crit}$, where $\rho_{crit} = 3H^2/8\pi G$ is the critical density and $H$ is the Hubble parameter. From this, one can rewrite Eq.(\ref{EqFgas}) as
\begin{equation}
\label{RhoObs}
\rho_{\rm{c}}^{obs}(z) = \rho_{\rm{b}}(z) \left[ \frac{\gamma K}{f_{gas}^{X-ray}(z)}\left(\frac{d_L^\mathrm{fid}(z)}{d_L(z)}\right)^{3/2} - 1 \right],
\end{equation}
and obtain the observational value of the dark matter density, $\rho_{\rm{c}}^{obs}$, where the baryonic matter density evolves as $\rho_{\rm{b}}(z) = \rho_{\rm{b,0}}(1 + z)^3$ and we have transformed $d_A(z)$ into $d_L(z)$ using the cosmic distance duality relation (CDDR), $d_A(z)=d_L(z)/(1+z)^{2}$~\cite{Etherington07,Ellis07}\footnote{In recent years, several analyses have observationally tested this relation and verified its validity within 2$\sigma$ (see e.g. \cite{Holanda:2012at} and references therein). A table with current constraints on the CDDR can be found in \cite{Holandajcap2016}.}. Note that Eq. (\ref{RhoObs}) furnishes a model-independent estimate of the dark matter density provided that independent measurements of the gas mass fraction and luminosity distance at the same redshift, and  baryonic density are available. In the next section we describe the observational data sets used to estimate $\rho_{\rm{c}}(z)$, $\Omega_{\rm{c,0}}$ and $\Omega_{\rm{M,0}}$.

\section{Observational Data} \label{data}

The galaxy cluster sample used in our analysis was obtained from {\it Chandra} archive and consists of hot, massive, morphologically relaxed systems in redshift range $0.078 \leq z \leq 1.063$ -- see Panel 1 (left). In this sample, the 40 $f_{gas}$ measurements were obtained from spherical shells at radii near $r_{2500}$, which significantly reduces the corresponding theoretical uncertainty in gas depletion from hydrodynamic simulations \cite{Mantz2014}. Moreover, the bias in the mass measurements from X-ray data, which arises when one assumes hydrostatic equilibrium, was calibrated by robust mass estimates for the target clusters from weak gravitational lensing \cite{apple}.

We also use the full SNe Ia sample from the Pantheon compilation~\cite{pan}. This sample contains 1048 SNe Ia in redshift range of $0.01 \leq z \leq 2.3$, including 276 SNe Ia ($0.03 \leq  z \leq 0.65$) discovered by the Pan-STARRS1 Medium Deep Survey and SNe Ia distance estimates from SDSS, SNLS, CfA { {(1-4)}}, CSP and  HST samples. The light curves have high quality and were obtained by using an improved SALT2 method \cite{JLA}.  

{  In order to obtain $\rho_{\rm{c}}^{obs}$ from Eq. (\ref{RhoObs}), we need pairs SNe Ia and galaxy clusters measurements at the same redshift. We reconstruct the evolution of the  luminosity distance with redshift in a non-parametric way using the Gaussian Processes (GP) method~\cite{gapp,gapp2}. We transform  the Pantheon data into $d_L(z)$ measurements using the expression $d_L(z) = 10^{(\bar{\mu}(z) - 25)/5}$ and add the error of the absolute magnitude to the covariance matrix of the SNe Ia. Finally, taking into account the full covariance matrix of the SNe (statistics and systematic), we apply GP to obtain a continuous reconstruction of the luminosity distance\footnote{{In order to apply the GP reconstruction method to the transformed apparent magnitude data into luminosity distance data, we assume that the resulted distribution of the $d_L$ data is Gaussian. We confirm the validity of this approximation by performing the  Kolmogorov-Smirnov test to each luminosity distance estimate.}}. }

\begin{figure*}[t]
\centering
\includegraphics[width=0.324\textwidth]{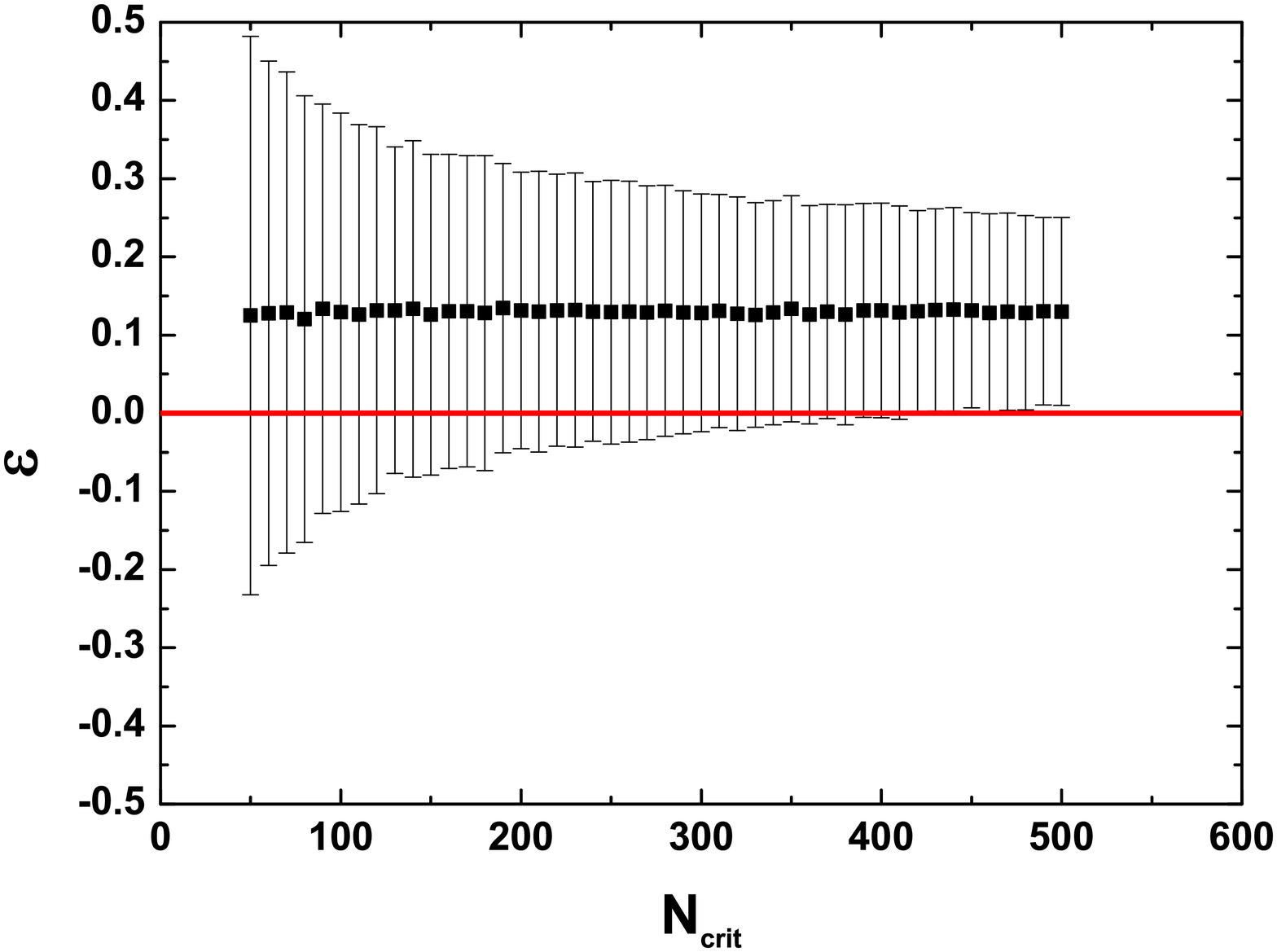}
\includegraphics[width=0.324\textwidth]{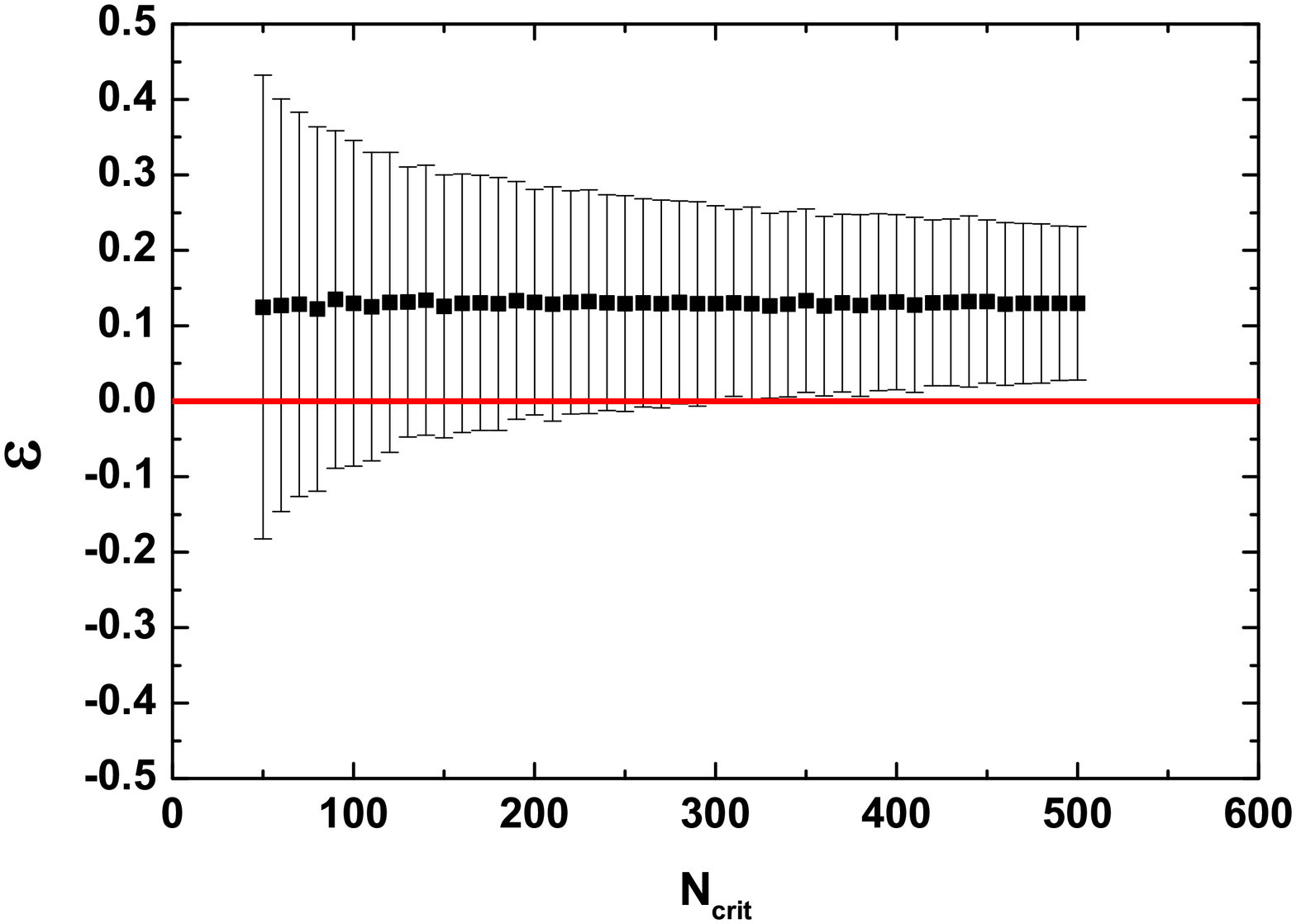}
\includegraphics[width=0.324\textwidth]{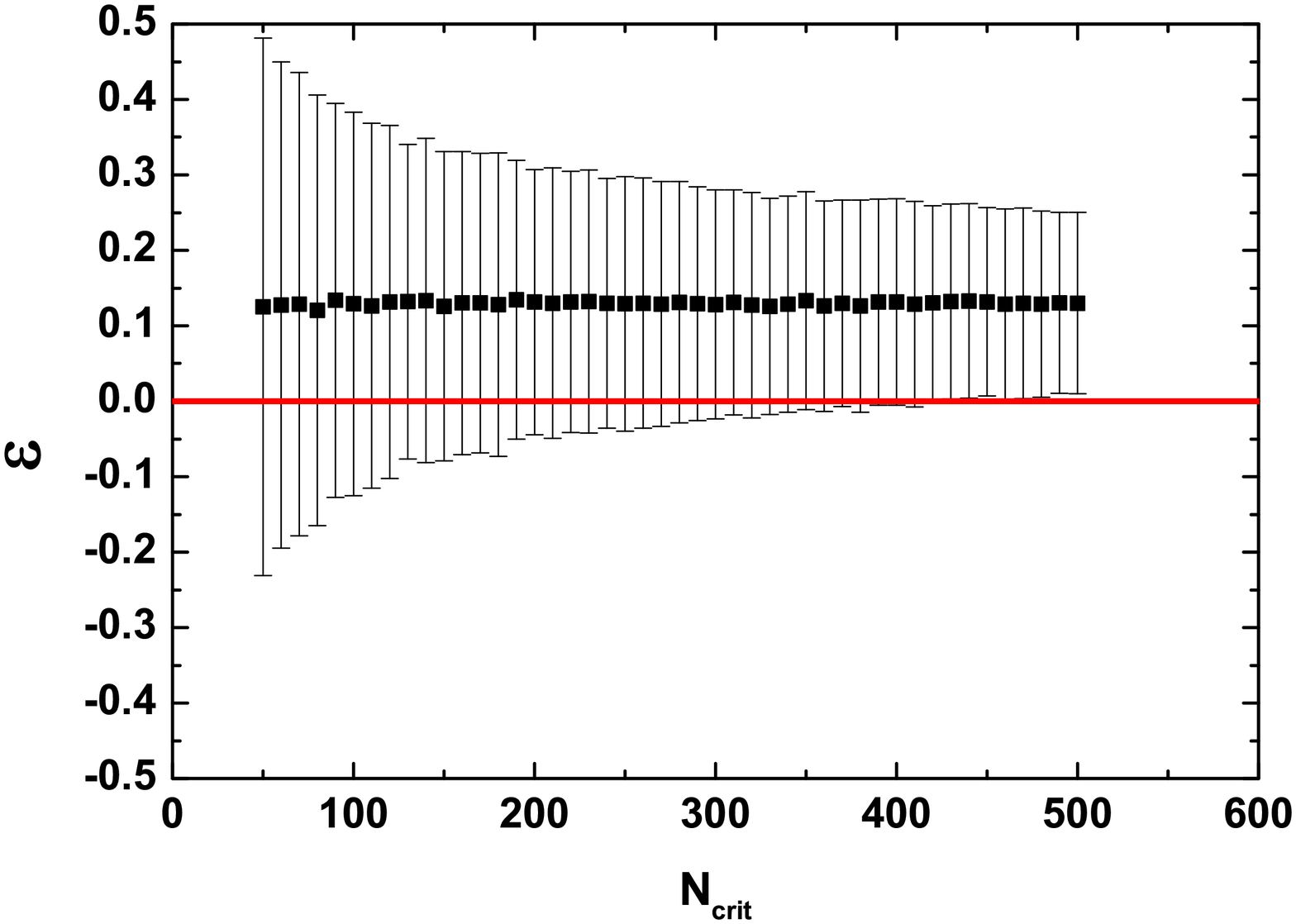}
\caption{The results of our simulations. {\it{Left)}} The variation of the $\epsilon$ parameter as the number of points varies in the simulations assuming the current observational uncertainties. {\it{Middle)}} The same as in previous panel but assuming that the errors of $f_{gas}$ are $50\%$ smaller than the current observational values. {\it{Right)}} The same as in previous panels but assuming that the errors of $D_L$ are $50\%$ smaller than the present observational values. The horizontal line in all figures corresponds to the standard case $\epsilon = 0$.}
\end{figure*}

Besides the $f_{gas}(z)$ and $d_L(z)$ measurements, we also use a recent determination of the baryonic matter density ($\rho_{{\rm{b,0}}}$), as reported in~\cite{Cooke2017}. The value, $\rho_{b,0} = 4.200 \pm 0.030\, (\times 10^{-31}$ g/cm$^3$), was obtained from a reanalysis of an absorption system located at $z = 2.52564$ towards the quasar Q1243+307 by using the echelle spectrograph on the Keck telescope.  Figure 2 shows the evolution of the dark matter density $\rho_{\rm{c}}^{obs}$ with the redshift obtained from the combination of data discussed above. The solid and dashed curves represent the best-fit values for the non-interacting and interacting frameworks which will be discussed in the next section.

\section{Analyses and results} 

\subsection{Non-interacting dark matter} \label{sec.NI}

Assuming a non-interacting dark matter, $\rho_{\rm{c}}(a) \propto a^{-3}$, we estimate the current dark matter density parameter from a $\chi^{2}$ statistics using the data set shown in Fig. 2. The corresponding likelihood, ${\cal{L}} \propto e^{- \chi^2}$,  as a function of $\rho_{\rm{c,0}}$ is shown in Panel 3 (left). From this analysis, we obtain $\rho_{\rm{c,0}} = 25.15 \pm 0.95\, (\times  10^{-31}$g/cm$^3$) at 1$\sigma$. Now, using the definition of the critical density we find $\Omega_{\rm{c,0}}h^2= 0.1338 \pm 0.0051$, which is $\simeq 2.6\sigma$ from the Planck estimate for the $\Lambda$CDM cosmology, $\Omega_{\rm{c,0}}h^2= 0.1202 \pm 0.0014$ (see Table II of \cite{Aghanim:2018eyx}, fifth column). By assuming a model-independent estimate of the Hubble parameter, $H_0 = 74.03 \pm 1.42$ (1$\sigma$), as reported in~\cite{Riess:2019cxk}, we find $\Omega_{\rm{c,0}} = 0.244 \pm 0.013$ (1$\sigma$) or still $\Omega_{\rm{M,0}} = 0.285 \pm 0.013$ (1$\sigma$).

\subsection{Interacting dark matter} \label{sec.I}

 As commented earlier, if the dark components have an additional, non-gravitational interaction between them, an energy exchange is also expected, which violates adiabaticity. The phenomenology of the interaction term has been extensively studied in the literature, and we refer the reader to~\cite{Cid:2018ugy} and references therein for a recent discussion. Here we adopt an empirical modification to the dark matter density evolution, which reproduces the functional form for a number of interacting models \cite{wang2,alcaniz2}:
\begin{equation} \label{idm}
\rho_{\rm{c}} = \rho_{\rm{c,0}}a^{-3 + \epsilon(a)}\;,
\end{equation} 
where $\epsilon(a)$ quantifies the rate of energy transfer. From a qualitative point of view, the above expression is jut telling us that due to the energy exchange between dark energy and dark matter, this latter component will necessarily dilute at a different rate when compared to its standard evolution, $\rho_m \propto a^{-3}$. For simplicity, we assume a constant value for interacting rate $\epsilon$. Note also that values of $\epsilon \ll 1$ are physically expected. 
 
Using the expression above,  we perform a $\chi^{2}$ statistics to estimate the parameters $\rho_{\rm{c,0}}$ and $\epsilon$. The result of this analysis is displayed in Panel 3 (right), which shows confidence contours (1$\sigma$ and 2$\sigma$) in the $\rho_{\rm{c,0}} -\epsilon$ plane. By marginalizing over $\rho_{\rm{c,0}}$ and $\epsilon$, we find, respectively (at 1$\sigma$):
$$
\epsilon = 0.13 \pm 0.235\;,
$$
$$
\rho_{\rm{c,0}} = 26.16 \pm 2.16\, (\times  10^{-31}\rm{g/cm^3}).
$$ 
From this latter value we also obtain $\Omega_{\rm{c,0}}h^2= 0.139 \pm 0.012$ and, assuming $H_0 = 74.03 \pm 1.42$~\cite{Riess:2019cxk}, we find $\Omega_{\rm{c,0}} =  0.254 \pm 0.023$ and $\Omega_{\rm{M,0}} = 0.295 \pm 0.023$  (1$\sigma$). {  For completeness, we also perform an additional analysis by varying the $K(z)$ and $\gamma(z)$ parameters within 1$\sigma$ (C.L.) of their  respective ranges. We find no significant difference from the above results.}

\section{Simulations} 

Clearly, the above constraints on the interaction parameter cannot discriminate between the interacting and non-interacting dark matter cases. In this section, we study the necessary improvements in number and accuracy of upcoming $f_{gas}$ and SNe Ia observations to detect a possible non-null value of the interacting parameter $\epsilon$.

We perform Monte Carlo simulations based on the current observational data discussed in Sec.~\ref{data}. The procedure followed  is:

\begin{itemize}

\item \textit{Fiducial model for the mean value:} We use the best fit parameters obtained in the previous section to define our fiducial model for the simulations. Then, a curve for $\rho_{\rm{c}}(z)$ is defined from this fiducial model, which for a specific redshift $z_{i}$, will provide  a fiducial value $\rho^{fid}_{\rm{c}}(z_i)$.

\item \textit{Fiducial curve for the errors:} By studying the error evolution for $f_{gas}$ and $d_L$ separately, we calculate the best fit curve for the error in each observable and state it as the initial fiducial curve for the errors. Then we exclude the points in a distance further than $1\sigma$ from this initial fiducial curve and recalculate the best fit curve. This final curve is assumed as the fiducial curve for the calculation of the errors. For a specific redshift $z_{i}$ we have a fiducial value for the error given by $\sigma^{fid}_{\rho_{\rm{c}}}(z_i)$. Also, the dispersion of the errors is calculated.

\item \textit{Redshift distribution of points:} We choose the number of points ($N$) that will compose our simulated data set and define the position of each point in equally spaced positions in redshift.

\item \textit{Simulation of a data set:} Based on the previous quantities, we simulate a data sample picking a random value with Gaussian distribution, in each redshift position $z_i$, given by $\cal{N}$ $(\rho^{fid}_{\rm{c}}(z_i),\sigma^{fid}_{\rho_{\rm{c}}}(z_i))$. Thus, by performing these simulations for the $N$ points previously chosen, we obtain one realization of a simulated data set.

\item \textit{Calculation of $\chi^2$ in each realization:} For the previous realization, we perform a $\chi^2$ analysis in order to obtain the best fit parameters of the interacting model (Eq.~\ref{idm}) for this realization. As we are specially interested in the evolution of the interaction under the changes of the observational errors, we take the best fit value of $\epsilon$.

\item \textit{Best fit for $\epsilon$:} We repeat the previous steps $10^4$ times and we obtain a mean value and standard deviation for the interaction parameter.

\item \textit{Variation on $N$:} We repeat the previous steps for different values of $N$ varying it in the range between $50 < N < 500$.

\item \textit{Variation on the observational errors:} We repeat the previous steps assuming a percentage fraction of the observational errors for $f_{gas}$ ($P_{\%}(\sigma_{f_{gas}})$) and separately for $d_L$ ($P_{\%}(\sigma_{d_L})$), varying the fraction of the observational errors in the range $50\%-100\%$.

\end{itemize}

The results of the simulations are shown in Fig. 4, where the vertical bars correspond to $1\sigma$ errors. Panel 4 (left) shows the evolution of the $\epsilon$ parameter inferred from the $\chi^2$ analysis as the number of points varies in the simulations whereas the Panel 4  (middle) shows the same as before, but assuming that the errors of $f_{gas}$ are $50\%$ smaller than the current observational values and the errors of $d_L$ are the same as the observational values discussed in Sec. \ref{data}. Finally, Panel (right) shows the same as before, but assuming that both the errors of $d_L$ and $f_{gas}$ are $50\%$ smaller than the current observational values.

As expected, Panel 4  (left) shows that the bigger the number of  pairs SNe Ia/galaxy clusters the smaller the predicted error bars. Thus, there is a number of pairs SNe Ia/galaxy clusters for which the error bar is equal to the distance between the mean value of $\epsilon$ and the standard value $\epsilon = 0$. We define this point as the critical number ($N_{crit}$) of pairs SNe Ia/galaxy clusters. For the number of observational data points bigger than $N_{crit}$,  an interaction within $68.3\%$ of confidence level is obtained. For the current observational errors, we find $N_{crit} = 419$. 

From Panels (middle) and (right) we obtain $N_{crit} = 301$ and $N_{crit} = 419$, respectively. Then, the results of the Fig. 4  (middle)  show that the variation on the $f_{gas}$ errors gives the major impact on the variation of $N_{crit}$. On the other hand, from Fig. 4 (right), the $N_{crit}$ value is approximately insensitive with respect to the improvements on the error bars of $d_L$. This last result is expected due to the fact that the uncertainties  of $\rho_{\rm{c}}(z)$ are mostly dominated by the $f_{gas}^{X-ray}$ errors.

\section{Conclusions}

In this paper we proposed and applied a cosmological model-independent approach to estimate the dark matter density and its time evolution. The main expression used in our analysis (Eq.~\ref{RhoObs}) relies solely on the validity of the CDDR. The analysis performed was divided into two parts. Firstly, we used a combination of $f_{gas}$, SNe Ia and $\rho_{b,0}$ observations to estimate the matter density parameter considering that the dark matter component is separately conserved. At 1$\sigma$, this analysis furnished $\Omega_{\rm{c,0}}h^2=0.1339 \pm 0.0051$, which is $\simeq 2.6\sigma$ from the Planck estimate for the $\Lambda$CDM cosmology. Then, we also considered an interacting dark matter component, with an evolution law of the type $\rho_{c} \propto a^{-3 + \epsilon}$. The analysis showed that the results are slightly modified with respect to the previous case, with the interacting parameter $\epsilon$ estimated to be close to zero. 

As the current constraints are not capable of confirming or ruling out an interaction in the dark sector, we used the present observational constraints as fiducial values and performed a series of Monte Carlo simulations in order to quantify the necessary improvements in number and accuracy of upcoming $f_{gas}$ and SNe Ia observations for a $\epsilon = 0$ detection. We then performed simulations of the $f_{gas}$ and $d_L$ data, increasing the number of points in the data sample between $50-500$ and defining the critical number of pairs SNe Ia/galaxy clusters, $N_{crit}$, as the number of observations for which a non-interacting model ($\epsilon = 0$) is discarded within $1 \sigma$. The current observational errors furnished $N_{crit} \simeq 419$ whereas considering $50\%$ of the original $f_{gas}$ errors, the critical number of points drops to $N_{crit} \simeq 301$. 

Finally, it is worth mentioning that the next generation of X-ray survey, such as eROSITA~\cite{erosita}, is expected to detect $\sim 100.000$ galaxy clusters, which may make possible a final answer about a non-gravitational interaction in the dark sector from the method discussed here, besides providing a model-independent estimate of the dark matter density parameter.

\section*{Acknowledgments}
RFLH acknowledges financial support from INCT-A and CNPq (No. 478524/2013-7, 303734/2014-0). RSG acknowledges support from PNPD/CAPES. JEG is supported by CNPq (PDJ
No. 155134/2018-3). JSA acknowledges support from CNPq (grant Nos. 310790/2014-0 and 400471/2014-0) and FAPERJ (grant No. E-26/203.024/2017).

\end{document}